\documentclass[aps,prd,showpacs, amssymb,superscriptaddress, notitlepage,floats,floatfix,nofootinbib]{revtex4-1}
\usepackage{graphicx}
\usepackage{amssymb}
\usepackage{amsmath}
\usepackage{amsfonts}

\usepackage{bm}
\usepackage{mathrsfs}
\usepackage[colorlinks=true]{hyperref}
\usepackage[all]{hypcap} 
\usepackage[utf8]{inputenc} 
\usepackage{mathptmx}      
\usepackage{color}
\usepackage{multirow}
\usepackage{palatino}
\usepackage{float}
\usepackage{journals}
\usepackage{soul}
\usepackage{calrsfs}
\DeclareMathAlphabet{\pazocal}{OMS}{zplm}{m}{n}

\begin{document}
\title{Wormhole Potentials and Throats from Quasi-Normal Modes}

\author{Sebastian H. V\"olkel }
\email{sebastian.voelkel@uni-tuebingen.de}
\affiliation{Theoretical Astrophysics, IAAT, University of T\"ubingen, Germany}
\author{Kostas D. Kokkotas}
\affiliation{Theoretical Astrophysics, IAAT, University of T\"ubingen, Germany}
\date{\today}
\begin{abstract}
Exotic compact objects refer to a wide class of black hole alternatives or effective models to describe phenomenologically quantum gravitational effects on the horizon scale. In this work we show how the knowledge of the quasi-normal mode spectrum of non-rotating wormhole models can be used to reconstruct the effective potential that appears in the perturbation equations. From this it is further possible to obtain the parameters that characterize the specific wormhole model, which in this paper was chosen to be the one by Damour and Solodukhin. We also address the question whether one can distinguish such type of wormholes from ultra compact stars, if only the quasi-normal mode spectrum is known. We have proven that this is not possible by using the trapped modes only, but requires additional information.
\par
The here presented inverse method is an extension of work that has previously been developed and applied to the oscillation spectra of ultra compact stars and gravastars. However, it is not limited to the study of exotic compact objects, but applicable to symmetric double barrier potentials that appear in one-dimensional wave equations. Therefore we think it can be of interest for other fields too.
\end{abstract}
\maketitle
%
\section{Introduction}
In this work we use the so-called trapped quasi-normal mode (QNM) spectrum of the Damour-Solodukhin (DS) wormhole \cite{PhysRevD.76.024016} to reconstruct the effective potential that appears in its wave equation for scalar perturbations. From this we can further recover the position of the wormhole throat, which in the DS wormhole is characterized by a single parameter $\lambda$. The present inverse method is an extension of recent work that has been done on the oscillation spectra of ultra compact stars \cite{paper1, paper2}.
As a proof of principle to demonstrate the inverse method, we apply it to the QNM spectra of the exact DS wormhole potential, as well as to the double P\"oschl-Teller (PT) potential, which can be used as an analytical approximation.
The QNM spectrum of both potentials have recently been studied \cite{Bueno:2017hyj}, where it was shown that the trapped QNM spectrum is encoded in the so-called ``echo'' signal of the object. Echo signals of this kind are expected to be emitted from a wide range of exotic compact objects (ECOs) \cite{2016PhRvL.116q1101C,2016PhRvD..94h4031C}, which could be different types of ultra compact stars, gravastars, wormholes or other exotic objects beyond general relativity, see \cite{Kokkotas:1995av,2000PhRvD..62j7504F,2004CQGra..21.1135V,Konoplya:2005et,PhysRevD.76.024016,PhysRevD.95.084034,2017arXiv170405789B,Barcelo2017,Berthiere:2017tms,Carballo-Rubio:2017tlh} for some examples. For a current overview about the topic we refer to \cite{Cardoso:2017njb,Cardoso:2017cqb}. These type of objects triggered a lot of attention recently, because there were claims that could be traced in gravitational wave data from LIGO \cite{2017PhRvD..96h2004A,2017arXiv170103485A,2017arXiv171206517C,Abedi:2018pst}. However, due to technical difficulties in the waveform modeling and data analysis, these claims are disputed \cite{2016arXiv161205625A,2017arXiv171209966W}. Although much effort has already been made to produce and model the expected signal \cite{Nakano:2017fvh,2017PhRvD..96h4002M,paper3,2017arXiv171002136G,Bueno:2017hyj,Correia:2018apm}, providing realistic waveforms remains a complicated problem due to the lack of concrete physically well motivated models. The existence of light primordial ECOs could also have drastic consequences for the nature of dark matter \cite{Raidal:2018eoo}.
\par
The inverse method we present is not limited or specially designed for the study of wormholes or other ECOs. We point out that it is a general inverse method to reconstruct a symmetric double barrier potential $V(x)$ that appears in the one-dimensional wave equation, where $E_n$ is the given spectrum. It is of approximate character because it is based on the ``inversion'' of the classical Bohr-Sommerfeld rule and the Gamow formula \cite{lieb2015studies,MR985100,1980AmJPh..48..432L,2006AmJPh..74..638G}, which can be derived from WKB theory. Besides presenting and applying the inverse method, \textit{we also address the uniqueness of the reconstruction and the question under what conditions one can distinguish such a wormhole potential from the typical potential of ultra compact stars.}
\par
This paper is organized as follows. First we present the inverse method being extended for this work in Sec. \ref{Method}. In Sec. \ref{Models} we discuss the DS wormhole model that is studied in this paper. The application of the inverse method along with the results for the reconstructed throat is shown in Secs. \ref{Reconstruction} and \ref{Throat}. We compare our results with the ones for ultra compact stars in Sec. \ref{Comparison to Ultra Compact Star Potentials}. The discussion and conclusions of our findings can be found in Secs. \ref{Discussion} and \ref{Conclusions}. We provide additional material in the Appendix \ref{Appendix}. Throughout the paper we use $G=c=\hbar=2m=1$.
\section{Inverse Spectrum Method} \label{Method}
\begin{figure}[H]
\centering
\includegraphics[width=8cm]{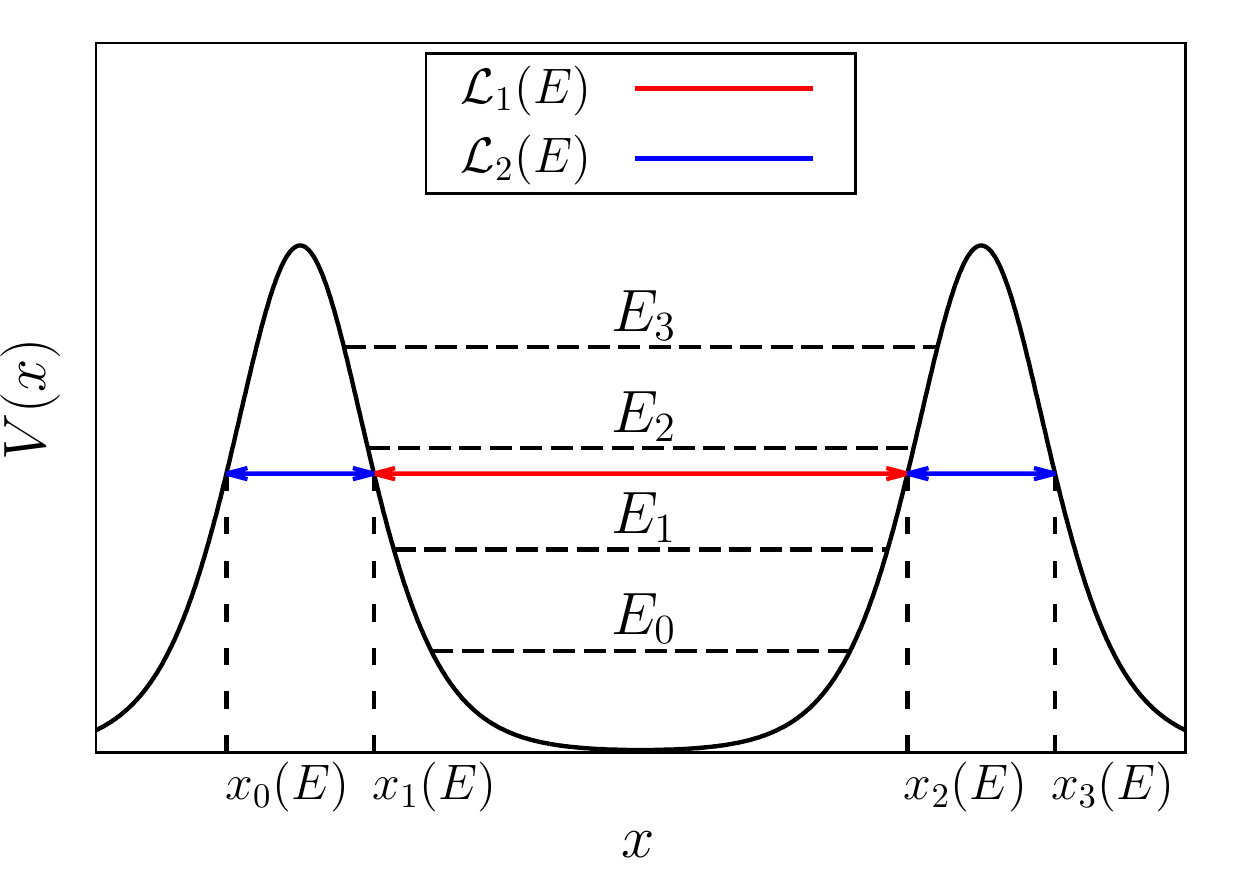}
\caption{Here we show a typical symmetric double barrier potential studied in this work. The red and blue arrows denote the width of the potential well $\pazocal{L}_1(E)$ and individual potential barrier $\pazocal{L}_2(E)$, respectively. The corresponding turning points at the value $E$ are indicated with broken lines. The real parts of the spectrum are shown as $E_i$. \label{potential_fig}}
\end{figure}

In this section we describe the method being used to reconstruct a symmetric double barrier potential of the type shown in Fig. \ref{potential_fig}. The potential is assumed to have one local minimum in the potential well and only two global maxima at the barrier. The method presented in this paper is an extension of the recently developed and applied method in the study of ultra compact stars, gravastars and mathematical physics \cite{paper2,paper4}. It is a generalization of inverse problems with pure potential wells and pure potential barriers \cite{lieb2015studies,MR985100,1980AmJPh..48..432L,2006AmJPh..74..638G}, which have been used in molecular spectroscopy and quantum algebra \cite{1991JPhA...24L.795B,1992JMP....33.2958B,PhysRevA.45.R6153}. The underlying idea is to invert a generalized Bohr-Sommerfeld rule \cite{1991PhLA..157..185P,2013waap.book.....K}, which can in our case be approximated with the classical Bohr-Sommerfeld rule and the Gamow formula. It is possible to use the inverted Bohr-Sommerfeld rule to reconstruct the width of the potential well $\pazocal{L}_1(E)$. The same can be done with the inverted Gamow formula, in combination with $\pazocal{L}_1(E)$, to reconstruct the width of the potential barrier $\pazocal{L}_2(E)$\footnote{$\pazocal{L}_2(E)$ corresponds to the width of one of the symmetric potential barriers, if the barriers are not symmetric the problem is more involved.}. In the subsequent sections we outline the extension of the inverse method and demonstrate how it is used. For more detailed discussions about this we refer to \cite{paper2,paper4,lieb2015studies,MR985100,1980AmJPh..48..432L,2006AmJPh..74..638G}.
\subsection{Outline of the Method}
The inverse method is applicable to the one-dimensional wave equation
\begin{align}
\frac{\text{d}^2}{\text{d}x^2}\psi(x) + \left(E_n-V(x) \right)\psi(x)=0,
\end{align}
where $V(x)$ is a symmetric double barrier potential and $E_n$ the corresponding spectrum. In most literature on this problem one usually finds $E_n$, and not the QNM notation $\omega_n^2$, as definition for the spectrum. We adopt this notation. The connection between the QNMs $\omega_n$ and $E_n$ is given by 
\begin{align}
E_n = E_{0n}+iE_{1n} = \Re\left(\omega_n^2\right)+ i \, \Im\left(\omega_n^2\right),
\end{align}
where $E_{0n}$ and $E_{1n}$ are the real and imaginary part of $E_n$. In this work we are interested in the trapped modes, also called quasi-stationary states, which correspond to the part of the spectrum related to potential well. For these modes one finds an exponentially small imaginary part $E_{1n}$, while the real part $E_{0n}$ is between the potential well minimum and the maxima of the barriers. Because of this property, one can expand the generalized Bohr-Sommerfeld rule \cite{1991PhLA..157..185P,2013waap.book.....K} and finds the classical Bohr-Sommerfeld rule describing the real part $E_{0n}$ from the potential well only
\begin{align}\label{cBS}
\int_{x_1}^{x_2} \sqrt{E_n-V(x)} \text{d}x = \pi \left(n+\frac{1}{2} \right),
\end{align}
where $(x_1, x_2)$ are the classical turning points in the potential well and $n \in \mathbb{N}_0$.
The imaginary part $E_{1n}$ then follows from the Gamow formula, which depends on the potential barrier and $E_{0n}$. Since the real part $E_{0n}$ can again be approximated with the classical Bohr-Sommerfeld rule, the result for the width of the potential well $\pazocal{L}_1(E)$ remains the same as in the case with one barrier
\begin{align}\label{Excursion}
\pazocal{L}_1(E) = x_2(E)-x_1(E) = \frac{\partial }{\partial E} I(E),
\end{align}
where $I(E)$ is the so-called inclusion and given by

\begin{align}\label{Inclusion}
I(E) = 2 \int_{E_\text{min}}^{E}\frac{n(E^\prime)+1/2}{\sqrt{E-E^\prime}} \mathrm{d}E^\prime.
\end{align}
From here on we denote $E$ as continuous real part of $E_{n}$ and $E_\text{min}$ is the minimum of the potential well. If the spectrum is not known analytically as function of $n$, one has to interpolate it. $E_\text{min}$ is not directly known from the spectrum, but can be approximated from the value where $n(E_\text{min})+1/2$ extrapolates to zero \cite{lieb2015studies,MR985100}.
\par
To generalize the method to the double barrier potential problem, one has to take into account the contribution of the second potential barrier to the imaginary part $E_{1n}$, as shown in \cite{2013waap.book.....K}. The Gamow formula now takes the form
\begin{align}
E_{1n} =& - \frac{1}{2}
\left(T_1(E) + T_2(E)\right) \left(\int_{x_1}^{x_2}\frac{1}{\sqrt{E_n-V(x)}} \text{d}x \right)^{-1} ,\label{E1n_1}
\end{align}
where $(x_0, x_1, x_2, x_3)$ are the four classical turning points for the quasi-stationary case and $T_{1,2}(E)$ are the semi-classical descriptions for the transmission through the potential barriers
\begin{align}
T_{1}(E) = \exp\left(2 i \int_{x_{0}}^{x_{1}} \sqrt{E_n-V(x)} \text{d}x\right),
\qquad
T_{2}(E) = \exp\left(2 i \int_{x_{2}}^{x_{3}} \sqrt{E_n-V(x)} \text{d}x\right).
\label{transmission}
\end{align}
The potentials being studied in this work are symmetric. In this special case the two integrals are identical and one finds an imaginary part that is two times larger than in the case with only one barrier
\begin{align}
E_{1n} =& -\left(\int_{x_1}^{x_2}\frac{1}{\sqrt{E_n-V(x)}} \text{d}x \right)^{-1} 
\exp\left(2 i \int_{x_0}^{x_1} \sqrt{E_n-V(x)} \text{d}x\right).\label{E1n_2}
\end{align}
Now one can apply the same inversion of the Gamow formula shown in \cite{1980AmJPh..48..432L,2006AmJPh..74..638G}. Note that the imaginary part $E_{1n}$ in eq. \eqref{E1n_2} requires the knowledge of the potential between the barriers between the classical turning points $(x_1,x_2)$, which is not known from the spectrum. However, in the appendix of \cite{paper2} it was shown that the knowledge of $\pazocal{L}_1(E)$, within the accuracy of the Bohr-Sommerfeld rule, is enough to uniquely describe $E_{1n}$ by choosing any valid potential constructed from $\pazocal{L}_1(E)$.
\par
One finds that the width of one of the potential barriers is given by
\begin{align}\label{widthbarrier}
 \pazocal{L}_2(E) = x_1(E)-x_0(E) =  \frac{1}{\pi} \int_{E}^{E_\text{max}} \frac{\left(\text{d}T(E^\prime)/\text{d}E^\prime \right)}{T(E^\prime) \sqrt{E^\prime -E}} \text{d} E^\prime,
\end{align}
where $E_\text{max}$ is the maximum of the potential barriers, which can be extrapolated from $T_{1,2}(E_\text{max})=1$ or provided otherwise. The transmission is obtained by combining eqs. \eqref{transmission} and \eqref{E1n_2}. If the spectrum is not known analytically, but as a discrete set of complex numbers, one has to interpolate between the states in order to obtain a discrete function for $T(E)$. From here one can already expect that the accuracy of the reconstruction improves for potentials with a large number of trapped modes compared to ones with a smaller number, see \cite{paper2,paper4} for discussions.
\par
With the reconstruction of $(\pazocal{L}_1, \pazocal{L}_2)$ one has two equations for four turning points $(x_0,x_1,x_2,x_3)$. Without additional assumptions that provide two turning points, one can not find a unique reconstruction of the turning points and from this the potential $V(x)$. It actually implies that infinitively many potentials, which must have the same $(\pazocal{L}_1, \pazocal{L}_2)$, yield exactly the same spectrum within the accuracy of the underlying WKB method. Although one might expect that this is a drawback of the non-exact WKB method, it simply says that the knowledge of the spectrum alone, is not enough for a unique reconstruction. A similar case has been coined with the famous question: \textit{Can one hear the shape of a drum?}\cite{10.2307/2313748,Gordon1992}.
\par
In the case of a symmetric potential, which was assumed to derive the above results, we can use the symmetry to reconstruct a unique potential. For a potential being symmetric around $x=0$, the turning points are given by\footnote{If the minimum of the potential well is not located at $x_\text{min}=0$, the derived turning points are all shifted by $x_\text{min}$. The spectrum is of course independent of such shifts.}
\begin{align}
x_0(E) = -\pazocal{L}_1(E)/2 - \pazocal{L}_2(E),\qquad
x_1(E) =-\pazocal{L}_1(E)/2 ,\qquad
x_2(E) = \pazocal{L}_1(E)/2,\qquad
x_3(E) = \pazocal{L}_1(E)/2 + \pazocal{L}_2(E).
\end{align}
The reconstructed potential is simply found by inverting the turning points $x_i(E)$ for $E$.
\section{Wormhole Models}\label{Models}
%
In this section we summarize the non-rotating wormhole model by Damour and Solodukhin \cite{PhysRevD.76.024016} and the double P\"oschl-Teller potential, which can be used as an approximation. The P\"oschl-Teller potential has a notable history in the study of QNMs of black holes and other compact objects, because it allows for explicit analytical calculations, we refer to \cite{1984PhRvL..52.1361F,1984PhRvD..30..295F,Konoplya:2005et,1742-6596-314-1-012074,Price:2017cjr,Bueno:2017hyj,Cardona:2017scd} for some examples.
\subsection{Damour-Solodukhin Wormhole}
Damour and Solodukhin describe their wormhole model as a black hole foil. Depending on a single parameter $\lambda$, it can describe the Schwarzschild space-time very well for large distances, but it becomes significantly different close to the Schwarzschild radius. The line element is given by
\begin{align}
\text{d}s^2 = -\left(g(r)+\lambda^2 \right) \text{d}t^2 + \frac{\text{d}r^2}{g(r)}+r^2 \left(\text{d}\theta^2+\sin^2(\theta) \text{d}\phi^2 \right),
\end{align}
with $g(r) = 1-2M/r$. To simplify the following relations we set $r/M \rightarrow r$ and for the later introduced tortoise coordinate $r^{*}/M \rightarrow r^{*}$.
\par
The most striking difference compared to the Schwarzschild case is the absence of an event horizon. It is replaced by a ``throat'' that connects two isometric, asymptotically flat regions. The absence of the event horizon changes the QNM spectrum drastically, although the exterior part of the space-time can be almost the same far away from the throat. Perturbations of the DS wormhole have been studied in \cite{Bueno:2017hyj}, thus we only summarize the important relations we need for the inverse problem and refer to the same paper for more details.
\par
In the case of scalar perturbations one finds that the effective potential of the DS wormhole is given by
\begin{align}
V_l(r) = \frac{l(l+1)f(r)}{r^2}+\frac{\left(f(r)g(r) \right)^{\prime}}{2r},
\end{align}
with $f(r) = 1-2/r$ and $g(r)=1-2(1+\lambda^2)/r$. The wave equation has the standard form
\begin{align}
\frac{\text{d}^2}{\text{d}{r^{*}}^2} \psi + \left(\omega_n^2 -V_l(r) \right) \psi = 0.
\end{align}
The tortoise coordinate $r^{*}$ can be derived explicitly 
\begin{align}
r^{*} = \pm  \left( \sqrt{\left(r-2 \right) \left(r-2(1+\lambda^2) \right)}+(2+\lambda^2)\cosh^{-1}\left[\frac{1}{\lambda^{2}}\left(r-2\right)-1 \right]\right).\label{r_tort}
\end{align}
To obtain a double barrier potential studied in this work, one has to choose small values of $\lambda$. For the relevant range of $\lambda$, the following simplification gives a good approximation
\begin{align}
r^{*} \approx \label{tort_app}
r+2\ln\left(\frac{r}{2}-1\right)+2\left(\ln\left(\frac{4}{\lambda^2}\right)-1\right).
\end{align}
In this case one can relate the position of the throat, described by $\lambda$, with the distance $L$ between the two maxima of the potential barriers
\begin{align}
L \approx 4\left(\ln\left(\frac{4}{\lambda^2} \right)-1 \right),\label{throat}
\end{align}
while the first term in eq. \eqref{tort_app} corresponds to the usual tortoise coordinate for the Schwarzschild BH.
\subsection{Double P\"oschl-Teller Potential}
%
The double P\"oschl-Teller potential has recently been used in the same work to approximate the QNM spectrum of the
DS wormhole \cite{Bueno:2017hyj}. For large separations between them one can treat the full potential either as two single P\"oschl-Teller potentials, defined piecewisely on each side, or as their sum, defined on the full range. This three parameter potential can be written as
\begin{align}
V_{\text{PT}}(x) = \frac{V_0}{\cosh^2\left(a\left(x+L/2 \right)\right)} + \frac{V_0}{\cosh^2\left(a\left(x-L/2 \right)\right)},
\end{align}
where $V_0$ is the maximum of the potential barrier(s), $L$ is the difference between the two barrier maxima, and $a$ scales the second derivative at the barrier maximum. All three parameters can easily be obtained from the full effective potential of a given model for a given $l$ and $\lambda$
\begin{align}
V_0\equiv V(r^{*}_\text{max}),
\qquad
a\equiv \sqrt{-\frac{1}{2}\frac{V^{\prime \prime}(r^{*}_\text{max})}{V(r^{*}_\text{max})}},
\qquad
L\equiv 2 r^{*}_\text{max}.
\end{align}
The parameters $(a, V_0)$ depend strongly on the value of $l$ in the effective potential, while $L$ is a measure for the position of the throat and related to $\lambda$. An implicit equation for the analytic solution of the QNM spectrum $\omega_n$ is provided in the appendix of \cite{Bueno:2017hyj}
\begin{align}\label{PT_wn}
e^{-i\omega_nL} = e^{-i\pi n} \frac{\Gamma\left(1+i\omega_n/a \right)\Gamma\left(\xi-i\omega_n/a \right)\Gamma\left(1-\xi-i\omega_n/a \right)}{\Gamma\left(1-i\omega_n/a\right)\Gamma\left(\xi \right)\Gamma\left(1-\xi \right)},
\end{align}
with $\xi \equiv \left(1\pm i \sqrt{4 V_0/a^2-1}\right)/2$.
%
\section{Reconstruction of the Potentials}\label{Reconstruction}
In this section we first briefly show the results of the inverse method applied to the QNM spectrum of the exact DS wormhole potential, as well as to the spectrum of the double P\"oschl-Teller potential. In the subsequent Sec. \ref{Throat} we show how the parameter $\lambda$, which appears in the wormhole metric and is related to the position of the throat, can be recovered. The extended discussion of our results can be found in Sec. \ref{Discussion}.
\subsection{Damour-Solodukhin Wormhole}\label{DS_REC}
\begin{figure}[H]
\centering
	\begin{minipage}{0.45 \textwidth}
	\includegraphics[width=1.0\textwidth]{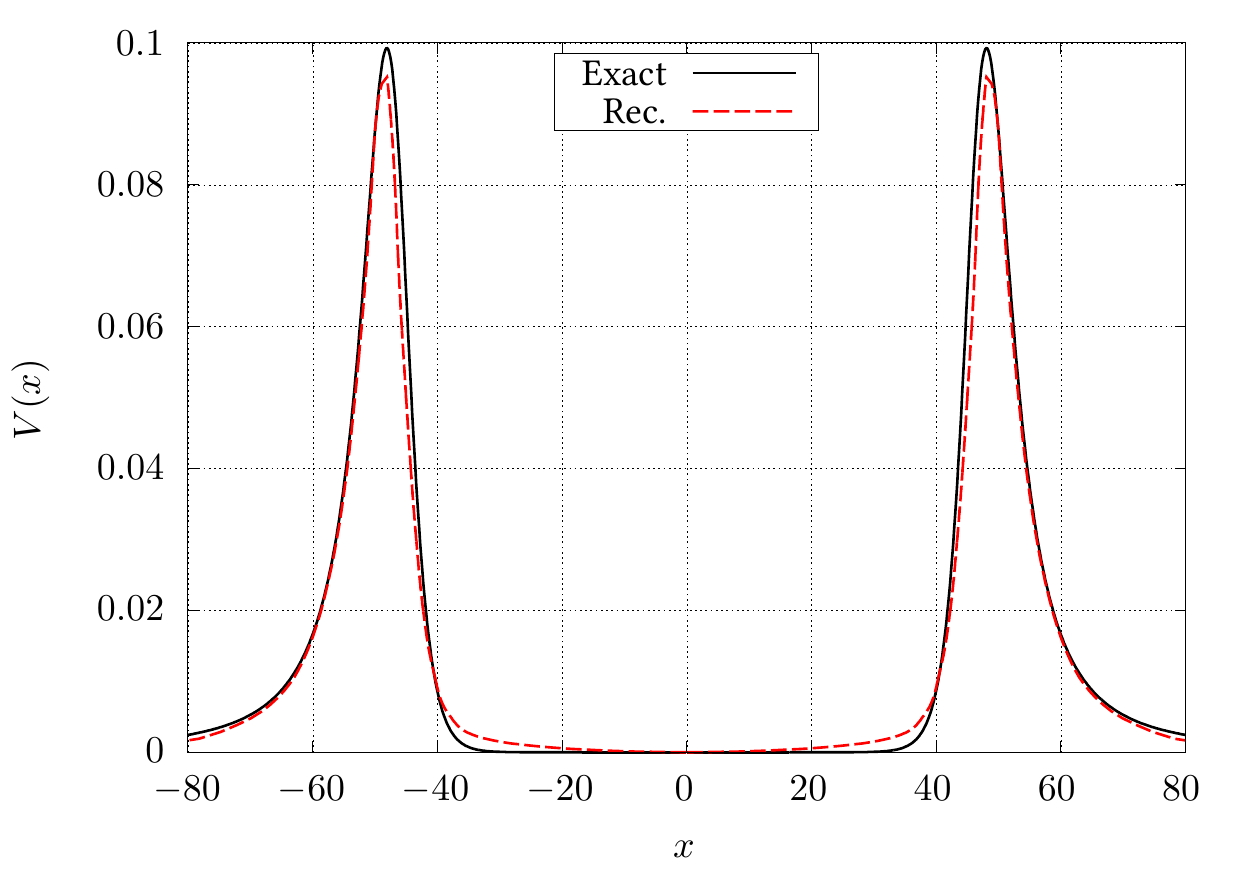}
	\end{minipage}
	\quad
	\begin{minipage}{0.45 \textwidth}
	\includegraphics[width=1.0\textwidth]{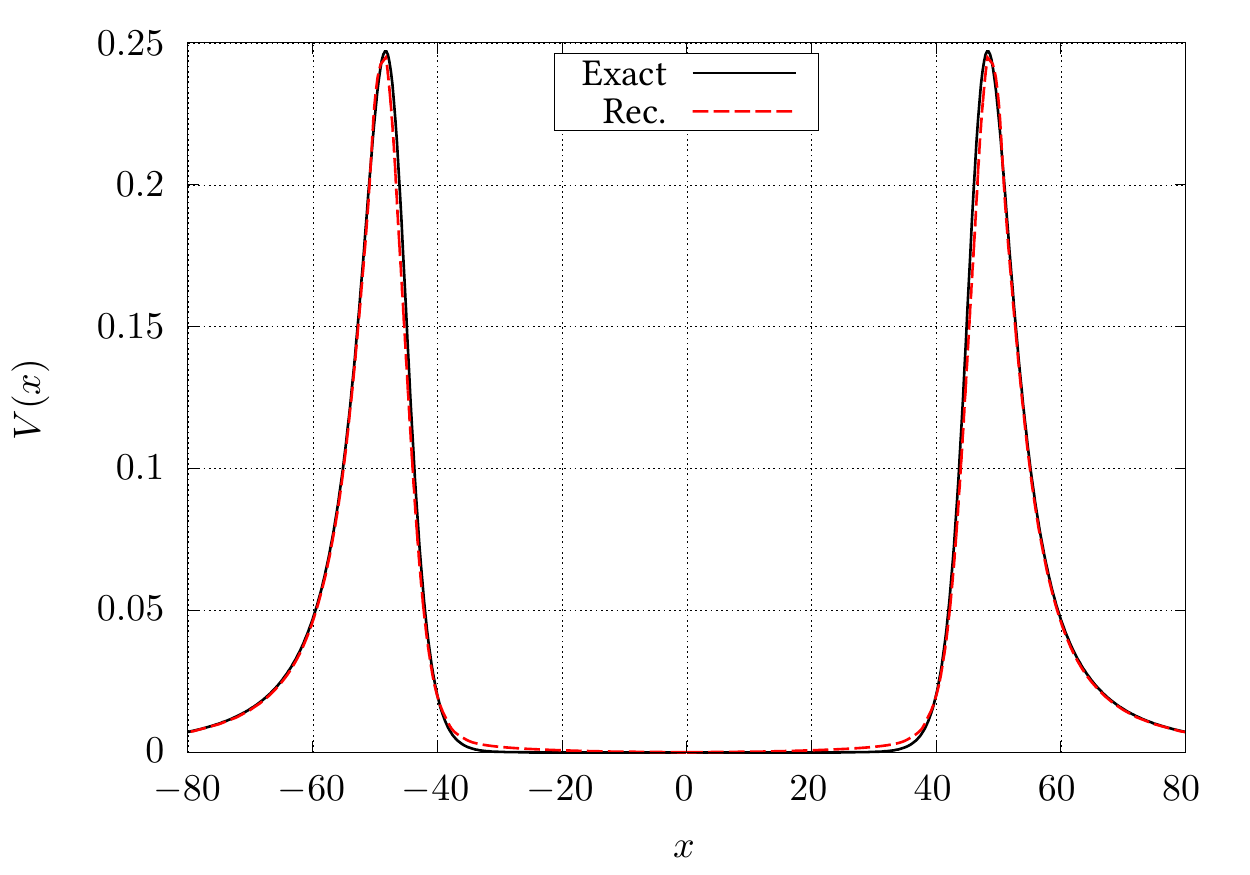}
	\end{minipage}
	\caption{In this figure we compare the exact DS potential (black solid) with our results for the reconstructed potential (red dashed) for $\lambda=10^{-5}$ and different values of $l$. \textbf{Left panel:} Here we show our result for $l=1$, there  are 9 trapped modes.
	\textbf{Right panel:} Here we show our result for $l=2$, there  are 14 trapped modes.\label{DS_REC_12}}
\end{figure}

Here we show the results for the reconstruction of the exact DS wormhole potential with the inverse method. The QNMs being used for the reconstruction are tabulated in Appendix \ref{appendix_tables} and have been produced with a modified code that was first presented in \cite{1994MNRAS.268.1015K}. The subsequent Fig. \ref{DS_REC_12} shows our result for the reconstructed potential for $\lambda=10^{-5}$ and $l=1,2$. Additional results for $l=3$ are provided in Appendix \ref{appendix_figures}.
\subsection{Double P\"oschl-Teller Potential}
Although the double P\"oschl-Teller potential is not a specific wormhole model, it is worth to apply the inverse method to this potential as well, to see how well one can reconstruct it from its trapped QNM spectrum. The QNMs being used for the reconstruction are tabulated in Appendix \ref{appendix_tables} and have been obtained by solving eq. \eqref{PT_wn} for $\omega_n$ numerically. Fig. \ref{PT_REC_12} shows our result for $\lambda=10^{-5}$ and $l=1,2$. The case of $l=3$ is provided in Appendix \ref{appendix_figures}. The panels are organized in the same way as for the DS potential in Sec. \ref{DS_REC}.
\begin{figure}[H]
\centering
	\begin{minipage}{0.45 \textwidth}
	\includegraphics[width=1.0\textwidth]{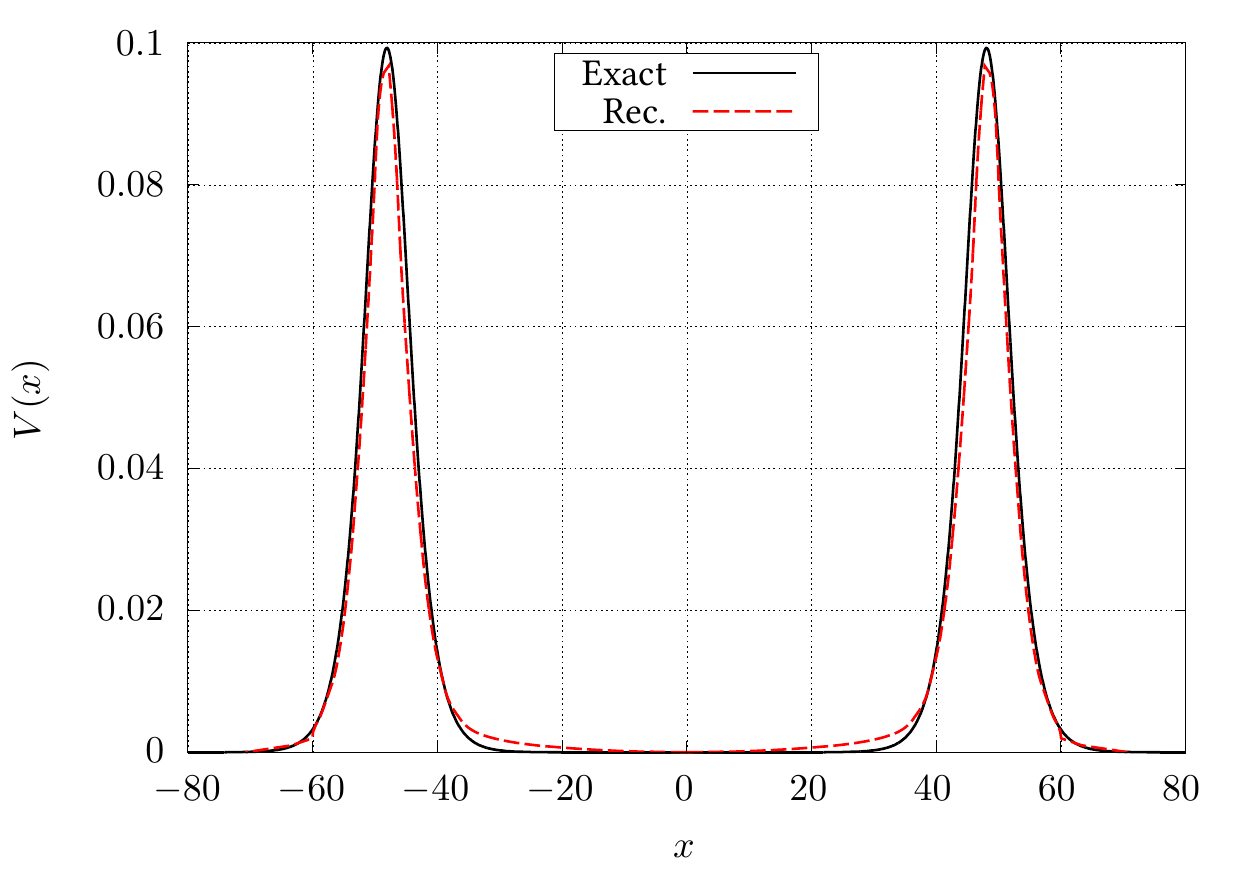}
	\end{minipage}
	\quad
	\begin{minipage}{0.45 \textwidth}
	\includegraphics[width=1.0\textwidth]{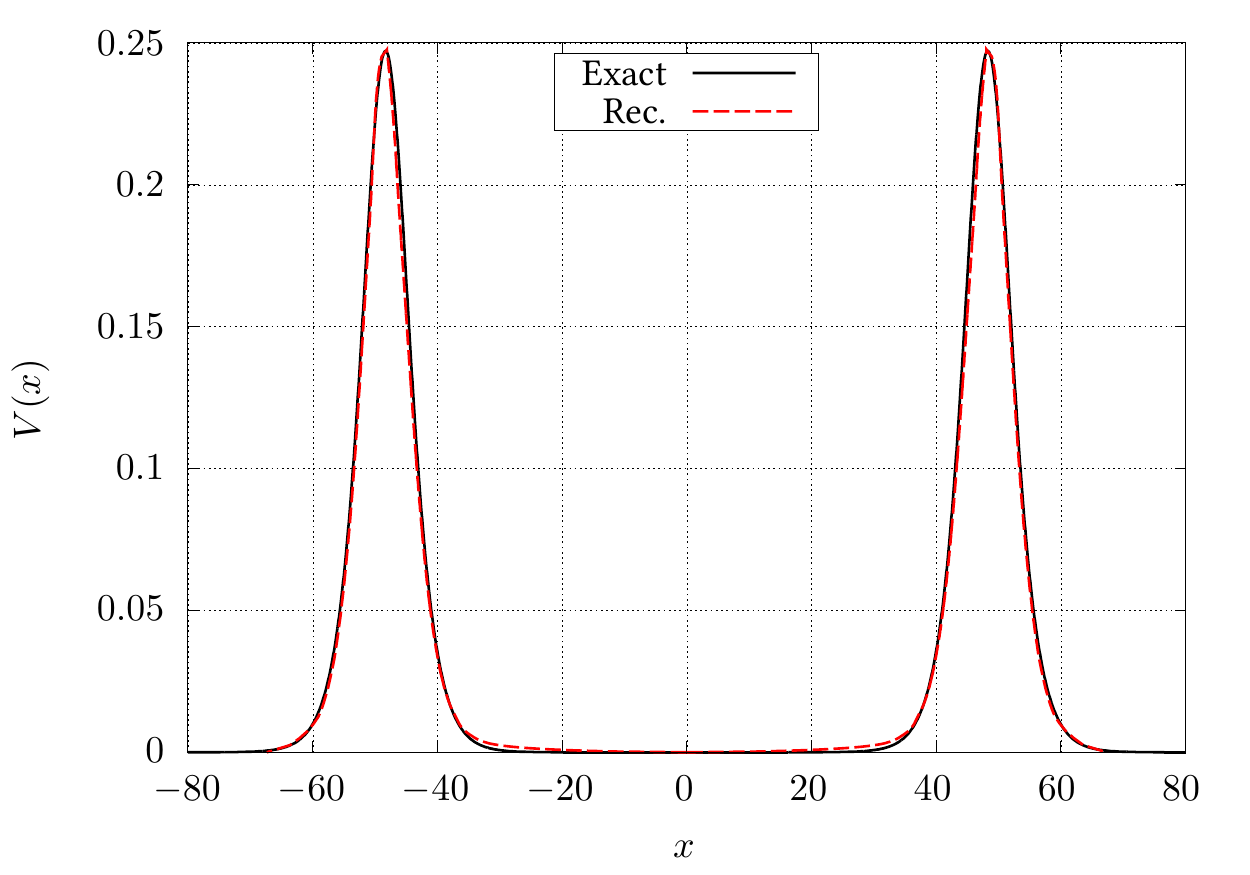}
	\end{minipage}
	\caption{In this figure we compare the double PT potential (black solid) with our results for the reconstructed potential (red dashed) for $\lambda=10^{-5}$ and different values of $l$. \textbf{Left panel:} Here we show our result for $l=1$, there  are 9 trapped modes.
	\textbf{Right panel:} Here we show our result for $l=2$, there  are 14 trapped modes.\label{PT_REC_12}}
\end{figure}
\section{Reconstructing the Throat}\label{Throat}
Once the DS potential has been reconstructed, we can recover the position of the throat being characterized by the parameter $\lambda$. So far we did not have to identify any specific property of the wormhole model for the inverse method\footnote{Besides the symmetric double barrier shape.}. However, deriving $\lambda$ from our results only makes sense, if one assumes that the underlying model for the wormhole is now the DS model.
\par
\begin{figure}[H]
\centering
\includegraphics[width=8cm]{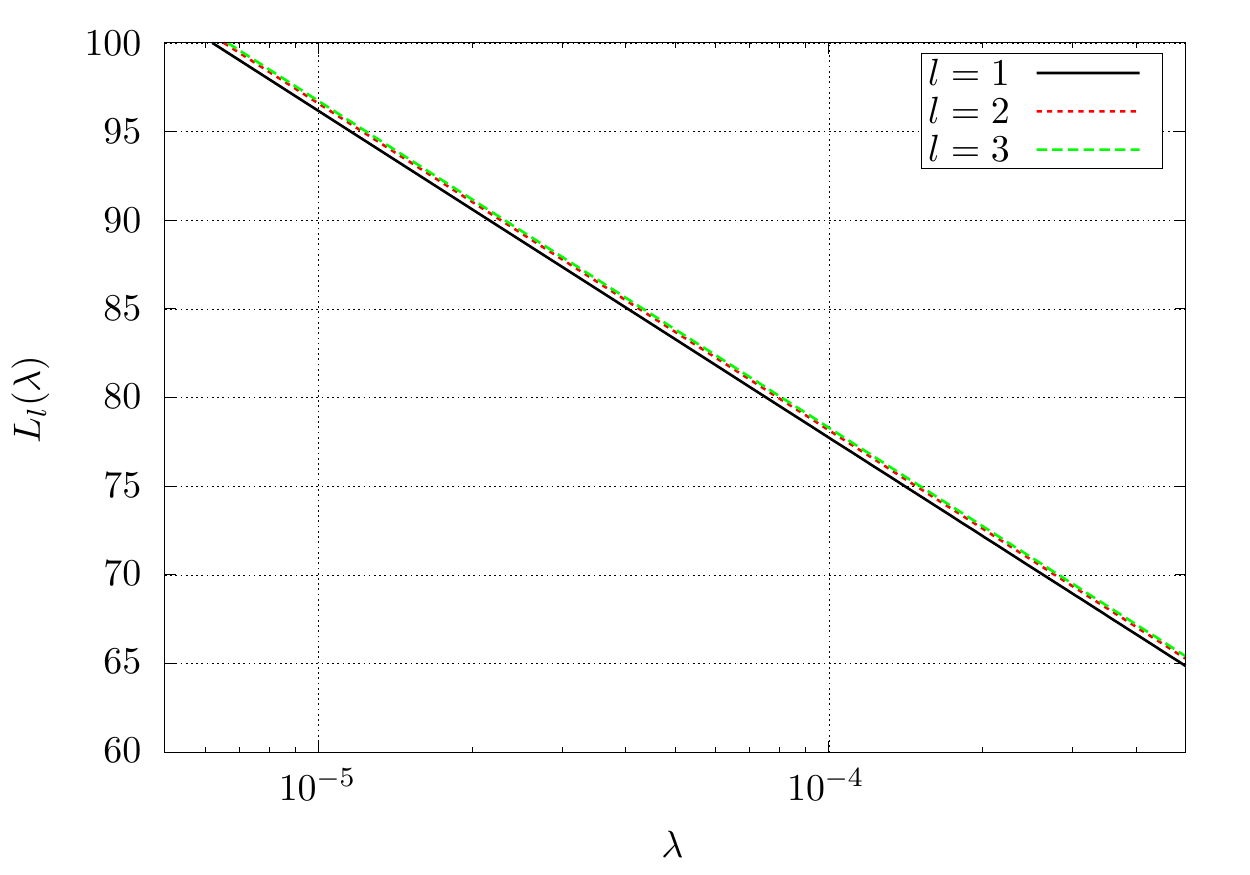}
\caption{Shown is the width of the potential well at $E_\text{max}/2$ as function of $\lambda$ for different values of $l$. Note that the $\lambda$-axis is in logarithmic scale.\label{throat_fig}}
\end{figure}
\begin{figure}[H]
\centering
\includegraphics[width=8cm]{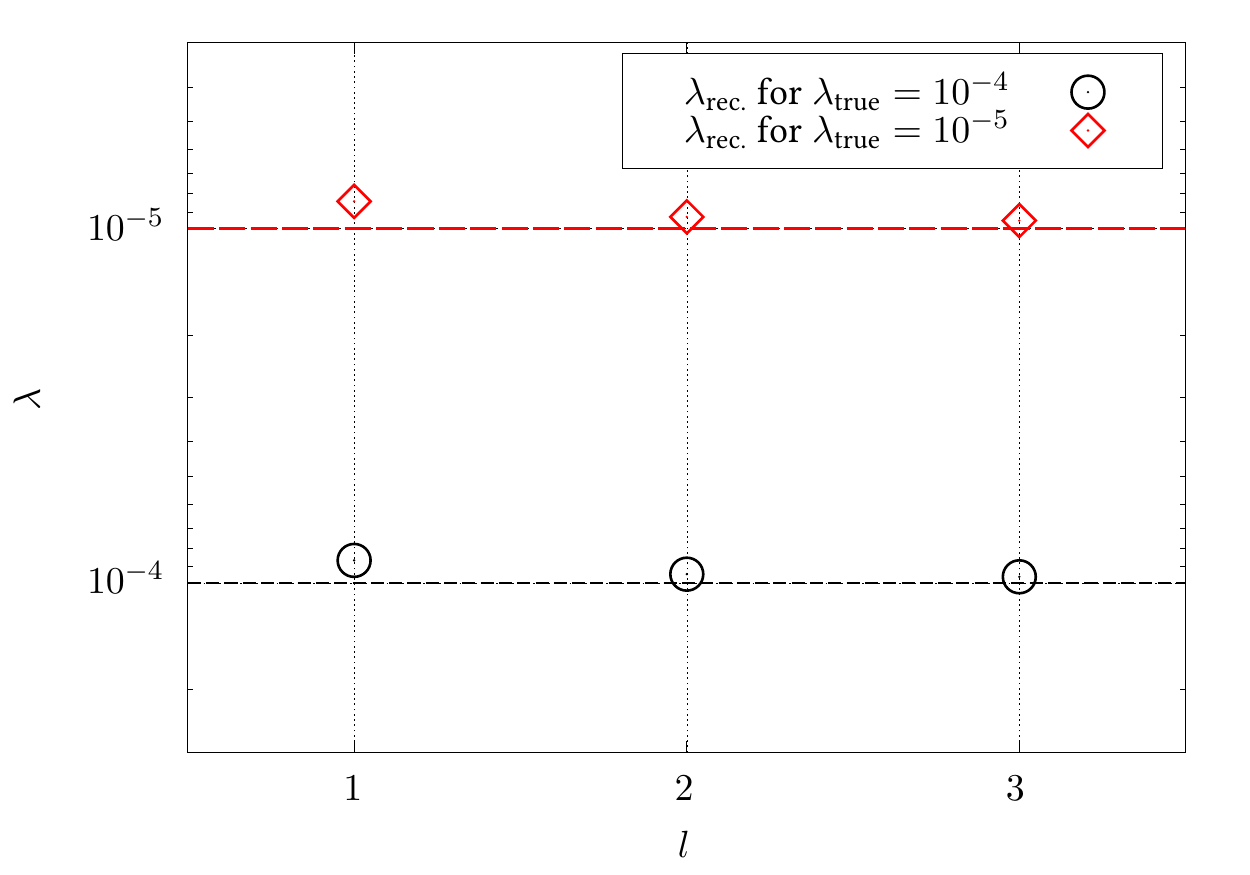}
\caption{Here we show the reconstructed values $\lambda_\text{rec.}$ (black circles and red diamonds) in the cases of $\lambda_\text{true}=10^{-4}$ (black dashed) and $\lambda_\text{true}=10^{-5}$ (red dashed). The y-axis is in a logarithmic scale, the tabulated values including the errors are provided in Sec. \ref{appendix_throat}.\label{LambdaRec}}
\end{figure}

To recover $\lambda$ in a simple way from the potential, we have chosen to use the distance $\pazocal{L}_1(E)$ evaluated at $E=E_\text{max}/2$, which is defined for every $l$ and $\lambda$. One can expect that the reconstructed potential is most precise around this region. For small values of $E$ the classical Bohr-Sommerfeld rule is less precise, while for values around the peak of the barriers, the Gamow formula becomes unreliable. For fixed values of $E=E_\text{max}/2$ and a given $l$, we define the following one-dimensional function
\begin{align}
L_l(\lambda) \equiv x_{2,l}\left(\lambda \right)-x_{1, l}\left( \lambda \right),
\end{align}
which for a given $l$ only depends on $\lambda$. Here $(x_{1, l}(\lambda),  x_{2,l}(\lambda))$ are the classical turning points for a given $l$ at $E=E_\text{max}/2$ as function of $\lambda$. We can use $L_l(\lambda)$ to bijectively relate the reconstructed width of the potential barrier with the corresponding value of $\lambda$. We show $L_l(\lambda)$ for the range of $\lambda$ being studied in this work in Fig. \ref{throat_fig}. The function depends only weakly on $l$. In order to recover $\lambda$ from the reconstructed potentials, one only has to solve the following equation for a given reconstructed width of the potential well $\pazocal{L}_1(E)$ (for given $l$)
\begin{align}
L_l(\lambda) = \pazocal{L}_1(E_\text{max}/2).\label{lambda_rel}
\end{align} 
This identification is a simple one which we use to demonstrate the precision one can expect from the method. In principle one could also reconstruct $\lambda$ with more refined methods, e.g. via $\chi^2$ fitting of the full functions for the reconstructed widths $(\pazocal{L}_1(E;\lambda), \pazocal{L}_2(E;\lambda))$ with the exact ones calculated directly from the DS potential as function of $\lambda$. We show the results for the reconstructed values of $\lambda_\text{rec.}$ for different parameters in Fig. \ref{LambdaRec}. Since the number of trapped modes increases with $l$, it is not surprising that the relative error becomes smaller with increasing $l$. We provide the tabulated values of the reconstructed values $\lambda_\text{rec.}$ along with the relative error in table \ref{Table_throat} in Appendix \ref{appendix_tables}. Concluding we want to note, as it is discussed in \cite{PhysRevD.76.024016}, that the time any perturbation takes to be reflected in the potential well, and thus excite the trapped QNMs, is increasing with smaller values of $\lambda$. Thus one should keep in mind that for too small values of $\lambda$, the trapped QNMs might be undetectable, since any observation can only cover a finite time interval.
%
\section{Comparison to Ultra Compact Star Potentials}\label{Comparison to Ultra Compact Star Potentials}
In this section we discuss and go beyond our results for the reconstructed wormhole potentials from Secs. \ref{Reconstruction} and \ref{Throat}. We compare them with similar results for ultra compact star potentials, reconstructed in \cite{paper2}, and ask whether one can distinguish both type of objects from the trapped QNM spectrum only. More precisely, we want to study the following two cases: first case, only the trapped QNMs are known; second case, additional information from the time evolution related to the scattering at the black hole like potential barrier is provided. To answer the question, we first discuss the general case of the two different type of potentials, before we embed it into the context of echoes from wormholes and ultra compact stars.
\subsubsection{Trapped Modes Equivalence}
The qualitative behavior of the trapped QNM spectrum for a double barrier potential is the same as for a single potential barrier next to a potential ``wall'', which is the situation for ultra compact stars. We call this kind of potential now to be of type I. The double barrier potentials are now called type II. The similarities between the two spectra is self-evident by looking at the Bohr-Sommerfeld rule eq. \eqref{cBS} and the Gamow formula for each problem\footnote{Only the imaginary part $E_{1n}$ is modified by a factor of two for symmetric barriers.}. Thus it seems interesting to ask the following question, for which the spectrum $\omega_n$ containing the trapped modes and potentially also the first scattered modes is known. \textit{Can one find two potentials, one of type I and one of type II, that admit the same trapped modes?} In the case of type II potentials studied in this work, we assume the potential to be symmetric around $x=0$. This assumption for the symmetry allows to reconstruct one unique potential. But without this symmetry, ``shifted'' or ``tilted'' potentials with the same widths $(\pazocal{L}_1(E), \pazocal{L}_2(E))$ would also have the same spectrum $\omega_n$. This is valid within the underlying first oder WKB theory. The situation for type I potentials is similar. Without the knowledge of one of the three turning point functions, one can find infinitely many different potentials with the same widths and therefore the same spectrum. However, in both cases this non-uniqueness is ``just'' within the same type of potential. \textit{But, is the same also true between the two different types of potentials? Can one find two potentials of type I and type II that describe the same spectrum?}
\par
To investigate this question, within the accuracy of the underlying WKB methods being used in this work, we can start from a given spectrum $\omega_n$ that contains all the trapped modes and potentially the first few scattered modes. Then we insert it into the two inverse methods. If we can construct pathology free widths $(\pazocal{L}_1(E), \pazocal{L}_2(E))$ in both cases, and from this provide at least one specific potential of each type, one can not distinguish the two different types of potentials from the knowledge of their trapped modes only. Again this is only valid within first order WKB theory, but can be seen as a strong hint that exact potentials, which are quite similar, exist as well.
\par
Note that by only looking at the full QNM spectrum (trapped modes and scattered modes), it is not obvious where exactly the trapped modes end and the scattered modes begin. Using the inverse methods, we can extrapolate the maximum of the potential barrier $E_\text{max}$ from the point where the semi-classical description of the transmission $T(E)$ extrapolates to $1$\footnote{The semi-classical description of the transmission $T(E)$ becomes less precise close to the top of the barrier, still it is a useful approximation to extrapolate the peak of the barrier.}. Since the Gamow formula for the imaginary part $E_{1n}$ differs by a factor of $2$ for the two different types of potentials, see Sec. \ref{Method}, one should expect slightly different values for the maximum $E_\text{max}$. From this it directly follows that one might find a slightly different definition of which modes of the spectrum still belong to the trapped modes and which are the first scattered modes.
\par
\begin{figure}[H]
\centering
	\begin{minipage}{5.5cm}
	\includegraphics[width=5.7cm]{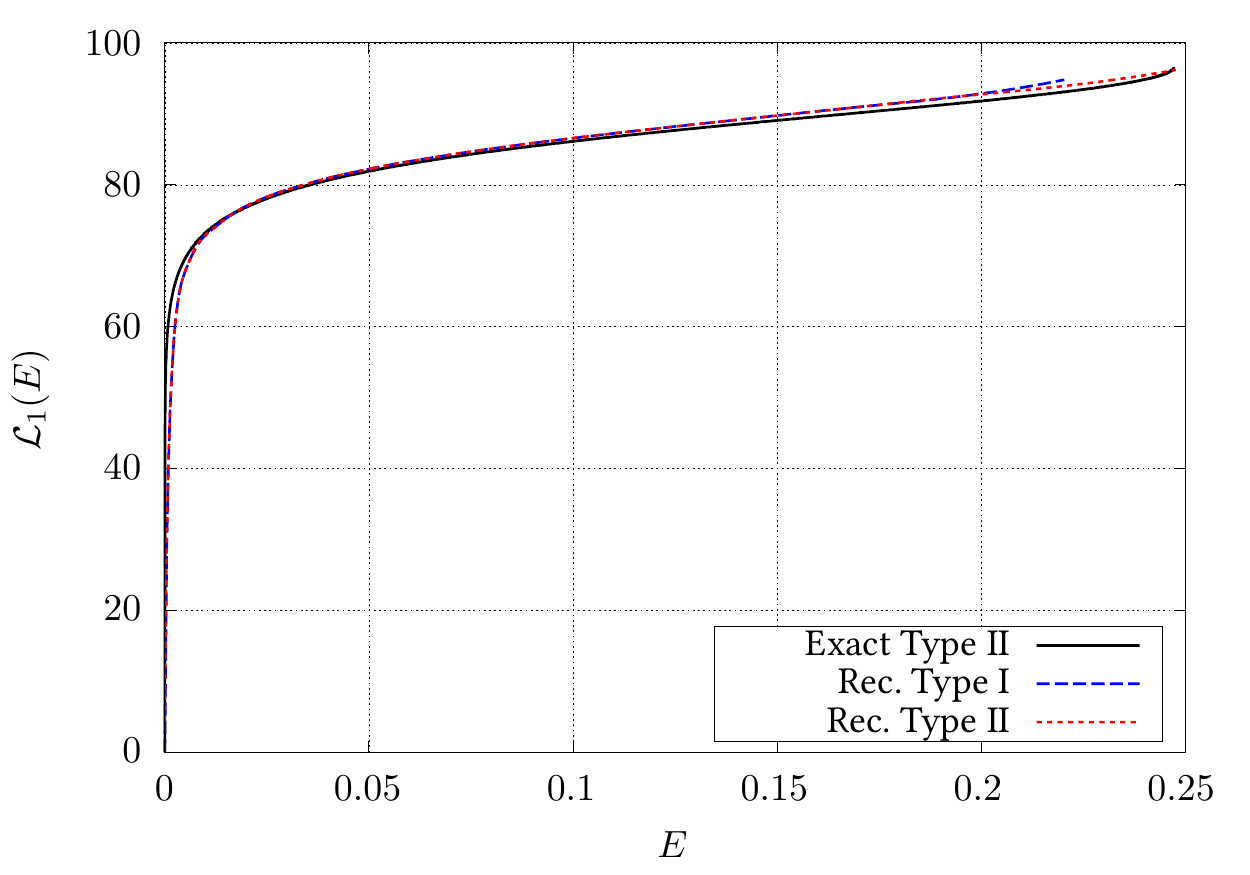}
	\end{minipage}
	\quad
	\begin{minipage}{5.5cm}
	\includegraphics[width=5.7cm]{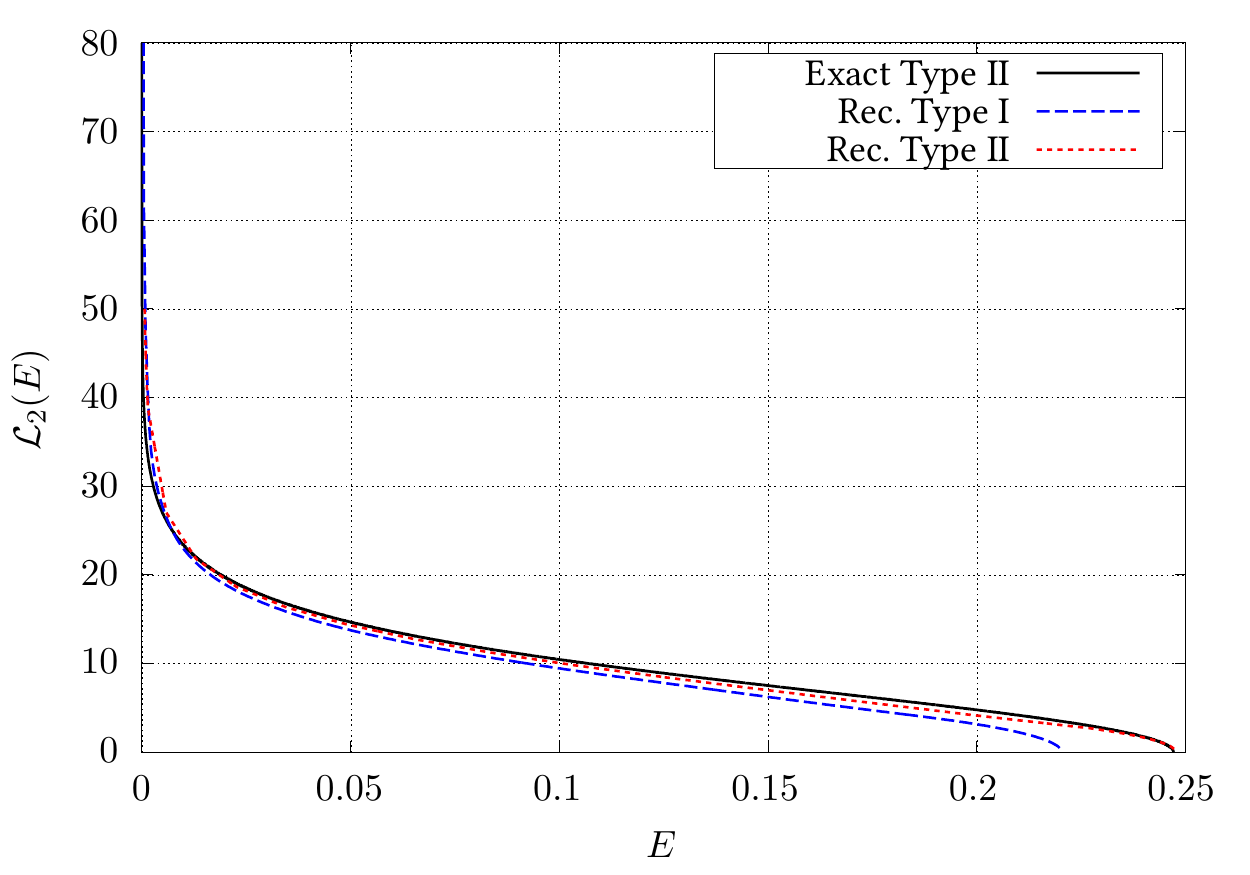}
	\end{minipage}
	\quad
	\begin{minipage}{5.5cm}
	\includegraphics[width=5.7cm]{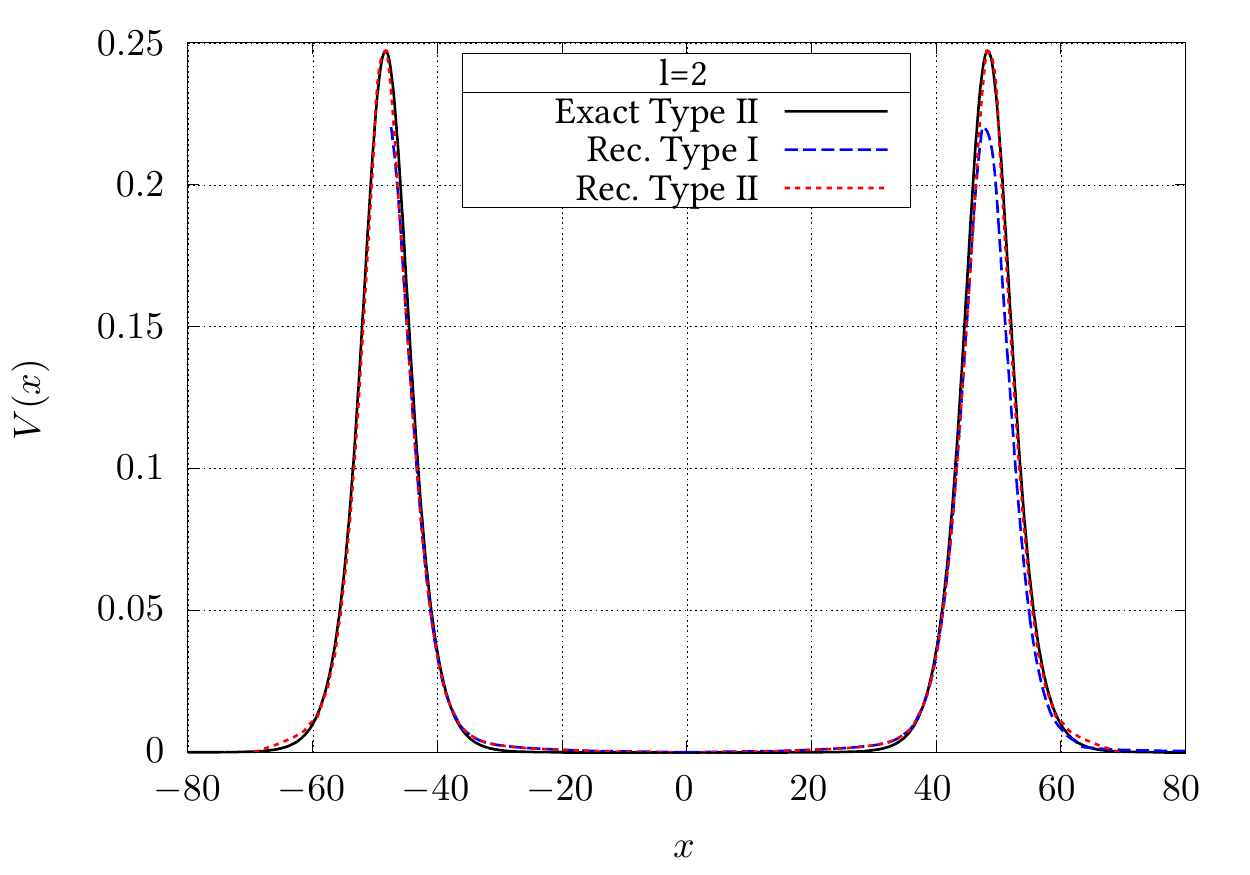}
	\end{minipage}
	\caption{Results for the \textit{trapped modes equivalent potentials}, one of potential type I, the other one of type II. The spectrum was obtained from the exact double P\"oschl-Teller potential for $\lambda=10^{-5}$ and $l=2$\label{WH_PT_l2_compared}, there are 13 trapped modes for type I and 14 trapped modes for type II. The exact  P\"oschl-Teller case is shown in solid black, the one of type I in dashed blue, and the one of type II in dashed red.
	\textbf{Left panel:} width of the bound region $\pazocal{L}_1(E)$. \textbf{Central panel:} width of the potential barrier $\pazocal{L}_2(E)$.\textbf{Right panel:} Exact potential and the two reconstructed trapped modes equivalent potentials of type I and II.}
\end{figure}

With all the previous thoughts in mind, we now show two \textit{trapped spectrum equivalent potentials} in Fig. \ref{WH_PT_l2_compared}. The order of the panels is the same as in the previous Sec. \ref{Reconstruction}. We can make several comments. First, due to the different Gamow formulas being used to approximate the transmission, the maximum of the potential barrier(s) differs slightly for both potentials. The maximum of the single barrier type is smaller than the one with two barriers. Second, since the real part is in both cases described by the same Bohr-Sommerfeld rule, it is not surprising that the width of the potential well $\pazocal{L}_1(E)$ is almost the same in both cases, it only differs by how far it is defined with respect to $E_\text{max}$ and deviates close to it. Third, the width of the potential barrier(s) $\pazocal{L}_2(E)$ differs close to the barrier maximum, but approaches each other with decreasing $E$. Note that the type I potential, shown in dashed blue in the right panel, goes only up to $E_\text{max}$. For type I potentials the left side continues to larger values, but it can not be reconstructed with the present method. For this one would need to generalize the inverse method for scattered modes, which is beyond the scope of this work.
To summarize this section, \textit{we have shown that two different type of potentials (type I and type II) can have the same trapped mode spectrum.} They differ in the width of the potential barrier(s) and have a slightly different value for the barrier maximum. It might happens that the total number of trapped modes is not the same, e.g. the first scattered mode of type I can be the last trapped mode of type II. All calculations and conclusions are only valid within the precision of the underlying first order WKB method. It would be interesting to verify whether this kind of equivalence is also true for higher scattered modes and how the exact equivalent potential looks like, but beyond the scope of this work. Since the Bohr-Sommerfeld rule and Gamow formula can be used to obtain quite precise spectra for both type of potentials, one should expect that the exact spectrum equivalent potentials are close to the ones we have presented in this work.
\subsubsection{Echoes from Wormholes vs. Ultra Compact Stars}
Now we can study the results of the previous section within the question whether one can distinguish a wormhole like object from an ultra compact star like object, if one only knows the trapped QNM spectrum contained in the expected echo signal. The QNMs in the echo signal are in principle all the trapped modes, but eventually also the first few scattered modes. It might also happen, that the last trapped mode coincides with the maximum of the potential barrier and thus could also be interpreted as first scattered mode. Therefore, it is in general not clear which part of the spectrum strictly belongs to the trapped modes and which describes the first scattered modes. Consequently, it might not be clear how many modes one has to use in the reconstruction. Choosing it by the value where the constructed transmission $T(E)$ (for both types of potentials) extrapolates to $1$, can lead to two different choices for the last trapped QNM.
\par 
Combining all the considerations from the previous section, \textit{one has to conclude that the two types of potentials can not be distinguished without additional information.} However, in the usual discussion of the expected echo signal, one knows that the first pulse of the echo signal is a direct reflection at the (first) potential barrier. This corresponds very well with the fundamental BH QNM \footnote{If the expected barrier is the same as for a BH. This has not to be the case in general, but the methods presented in \cite{1985ApJ...291L..33S,1984PhRvL..52.1361F,1984PhLA..100..231B,1987PhRvD..35.3621I} are valid for any type of potential barrier that can be approximated around the maximum.} that can be approximated with the Schutz-Will formula \cite{1985ApJ...291L..33S,1987PhRvD..35.3621I} or alternative descriptions \cite{1984PhRvL..52.1361F,1984PhLA..100..231B,1984PhRvD..30..295F}
\begin{align}
\omega_n^2 = V_\text{max}-i\left(n+\frac{1}{2} \right) \sqrt{-2V^{\prime \prime}_\text{max}},
\end{align}
where $V_\text{max}$ and $V^{\prime \prime}_\text{max}$ are the values of the potential barrier at the maximum and its corresponding second derivative. The most dominant mode is the fundamental one with $n=0$.
None of the reflected modes is part of the exact QNM spectrum of the ECO, but has to be understood as additional information related to the scattering at the potential barrier. With this additional information one can approximate the maximum of the potential barrier and its second derivative in an independent way and check which of the two reconstructed trapped mode equivalent potentials is the right one. However, precise knowledge of the reflected mode would be required to exclude one of the two potentials.
\par
The situation becomes even more clear, if one allows for additional physical assumptions. For example, in many ECO models, the modifications of the space-time become only important very close to the corresponding Schwarzschild radius, which can be at much smaller distances than the position of the barrier maximum. Thus one would expect most part of the barrier to be similar to the Schwarzschild case and can check which of the two reconstructed barriers fit.
\subsubsection{Wormholes vs. Black Holes}
Another fundamental question is whether black holes and wormholes (or other ECOs) can in general be distinguished from each other. It has been addressed several times in the literature, see \cite{2007CQGra..24.4191C,PhysRevD.76.124015,2016PhRvL.116q1101C,2016PhRvD..94h4016C,2016PhRvD..94h4031C,Konoplya:2016hmd,PhysRevD.95.104011} for some examples and the references given in \cite{Cardoso:2017njb,Cardoso:2017cqb}. Some of the findings might seem to be contradicting on the first view, but this is more due to the different types of wormholes being studied. The situation becomes rather clear in cases where an effective perturbation potential can be shown. Wormholes of the DS type, within the parameter range studied in this work, have an effective perturbation potential consisting of a potential well between two potential barriers. It is evident that the QNM spectrum is drastically different compared to a Schwarzschild black hole, whose perturbation potential is a single potential barrier. Therefore one can distinguish these objects from the late time behavior of emitted perturbations. However, there are also other wormhole models like the Bronnikov-Ellis wormhole \cite{Bronnikov:1973fh,Ellis:1973yv}, where the effective potential is a single potential barrier, which indeed can mimic the QNM spectrum and late time behavior of regular black holes \cite{Konoplya:2016hmd,PhysRevD.95.104011}.
%
\section{Discussion}\label{Discussion}
%
In this section we discuss the accuracy of the inverse method from the results we obtained in the previous sections and give a brief outlook regarding actual observations using gravitational wave detectors.
\subsection{Accuracy}
The accuracy is similar to the case of ultra compact stars \cite{paper2}. For the reconstructed widths $(\pazocal{L}_1(E),\pazocal{L}_2(E))$ one has to take into account the following aspects. First, the inverse method is a result of the WKB method and therefore \textit{intrinsically} approximate. Second, the total number of trapped modes. Closely related to this is the expectation that the WKB method can become very precise for high values of $n$, but less precise for small $n$\footnote{For the bound states in a potential well described by the Bohr-Sommerfeld rule.}. Something similar is true for the semi-classical description of the transmission. It can be quite precise for energy values far below the barrier maximum but becomes less precise if the energy approaches the value of the maximum. We thus observe that the overall accuracy of the reconstructed widths $(\pazocal{L}_1(E),\pazocal{L}_2(E))$ increases with the total number of trapped modes that exist in the potential, but has to be treated with caution close to the minimum of the potential well and close to the maximum of the barrier. Indeed, this behavior has been found in Sec. \ref{Reconstruction}. In cases where the widths are reconstructed without any pathologies\footnote{E.g. not being strictly increasing/decreasing functions of $E$.}, it can still happen that the reconstructed potential is unphysical due to ``overhanging cliffs'', which can come from the very large values of $\pazocal{L}_2(E)$ for small values of $E$ around $E_\text{min}$. For an extended discussion of this effect we refer to \cite{paper2,paper4}.
\par
So far it was assumed that the trapped QNM spectrum is known with pristine accuracy. Unfortunately, any real observation would come with an error which would also influence the reconstruction. Also, by analyzing the time-evolution of echoes, one finds that only the highest trapped QNMs are clearly visible, it will be more difficult to detect the lower ones. However, due to the approximative WKB method and the non-trivial integral equations that have to be solved, it is not obvious how exactly the final result is affected by initial errors on the spectrum. For small errors, which are also always present if one uses numerically obtained spectra, this effect does not play an important role. This view is supported by comparing the reconstructed results from the trapped QNM spectrum of the DS wormhole and the corresponding double P\"oschl-Teller potential. In both cases the reconstructed potentials look quite similar. Nevertheless, one essential part of the reconstruction is the inter-/extrapolation being used to define the continuous spectrum $n(E)$ and transmission $T(E)$. These functions can not be arbitrary but must fulfill several conditions to describe a physical system. In the semi-classical description they must be strictly monotone increasing functions of $E$. As soon as these conditions are not fulfilled, the reconstructed widths can become unphysical, which is a clear sign that the spectrum is erroneous, not originating from this type of potential or the inter-/extrapolation insufficient. A detailed study for cases where the errors are not negligible is a non-trivial task that remains to be done in the future.
\subsection{Gravitational Wave Detections}
Our last discussion is on how well current and next generation gravitational wave detectors could be used to recover the trapped QNM spectrum from the expected echo signal of exotic compact objects. This question has been addressed in recent work \cite{paper3}, where a parameter estimation for phenomenological echo templates has been done. The type of objects being studied there are ultra compact stars, not wormholes, but the phenomenological templates that describe the echo signals are similar in both cases \cite{2016PhRvL.116q1101C,2016PhRvD..94h4031C}. Therefore one can also apply the key results from there to the wormhole case and state that current and next generation gravitational wave detectors are in principle able to detect a few of the trapped QNMs, with non-negligible errors, but probably not all. It will be more complicated to detect the lower trapped modes than the higher ones. Because of this, but also due to the lack of concrete physical models that additionally also take into account rotational effects and deviations from axial symmetry, much more work has to be done, if one wants to make realistic statements about ECOs.
%
\section{Conclusions}\label{Conclusions}
Wormholes and ultra compact stars, usually called exotic compact objects, have drawn a lot of attention recently due to claims that new kind of signatures have been found in LIGO data \cite{2017PhRvD..96h2004A,2017arXiv170103485A,2017arXiv171206517C,Abedi:2018pst}. These findings are disputed \cite{2016arXiv161205625A,2017arXiv171209966W}, but still call for a more detailed study of the postmerger spectra. In this work we have successfully extended an inverse spectrum method, which was recently developed to reconstruct the effective perturbation potentials of ultra compact stars \cite{paper2}. The inverse method has been extended to reconstruct symmetric double barrier potentials from the knowledge of the trapped QNM spectrum. We applied the method to reconstruct the effective potential for scalar perturbations of the DS wormhole, which is characterized by symmetric double barriers. Starting from the trapped QNM spectrum of the exact DS wormhole and the double P\"oschl-Teller potential approximation, we were able to reconstruct both potentials for different parameters. In the subsequent step we assumed that the underlying wormhole model is the one by Damour and Solodukhin and recovered the $\lambda$ parameter, which describes the position of the throat, within a few percent.
\par
We also addressed the question whether it is possible to distinguish in principle the double barrier potential of some wormhole models from the potential well with a single barrier that appears for ultra compact stars and gravastars, by using only the trapped modes of the spectrum. To address the question we were able to explicitly reconstruct one potential of each type from the same QNM spectrum. Both potentials have the same trapped QNM spectrum within the accuracy of the underlying WKB theory, and are therefore indistinguishable. However, using that the first pulse of the expected echo signal is independent and additional information being related to the shape of the potential barrier, we argue that one can in principle distinguish them. Note that this additional information is not related to the QNM spectrum of the object, but follows from the time evolution of the scattering at the potential barrier.
\par
The inverse method is approximate since it is based on WKB theory, but has the advantage that it should be possible to extend it to different classes of potentials in the future and to include higher order correction terms for more precise reconstructions. We want to close our conclusions by emphasizing that the presented method is not limited to the study of exotic compact objects, but is applicable to reconstruct symmetric double barrier potentials in one-dimensional wave equations from the knowledge of the quasi-stationary states. Thus it should potentially find applications in other fields as well.
\acknowledgments
The authors would like to thank Roman Konoplya and Daniela D. Doneva for useful discussions and are very thankful for the constructive comments of the anonymous referees that improved the final version of this work. SV is indebted to the Baden-W\"urttemberg Stiftung for the financial support of this research project by the Eliteprogramme for Postdocs.
%
\bibliography{literatur1}
%
%
\section{Appendix}\label{Appendix}
%
In this appendix we provide additional results from the reconstruction of the potentials in Appendix \ref{appendix_figures} and the tabulated QNMs being used for the reconstruction in Appendix \ref{appendix_tables}. The tabulated values for the reconstruction of the throat parameter $\lambda$, including relative errors, are shown in Appendix \ref{appendix_throat}.
%
\subsection{Additional Figures}\label{appendix_figures}
\begin{figure}[H]
\centering
	\begin{minipage}{0.45 \textwidth}
	\includegraphics[width=1.0\textwidth]{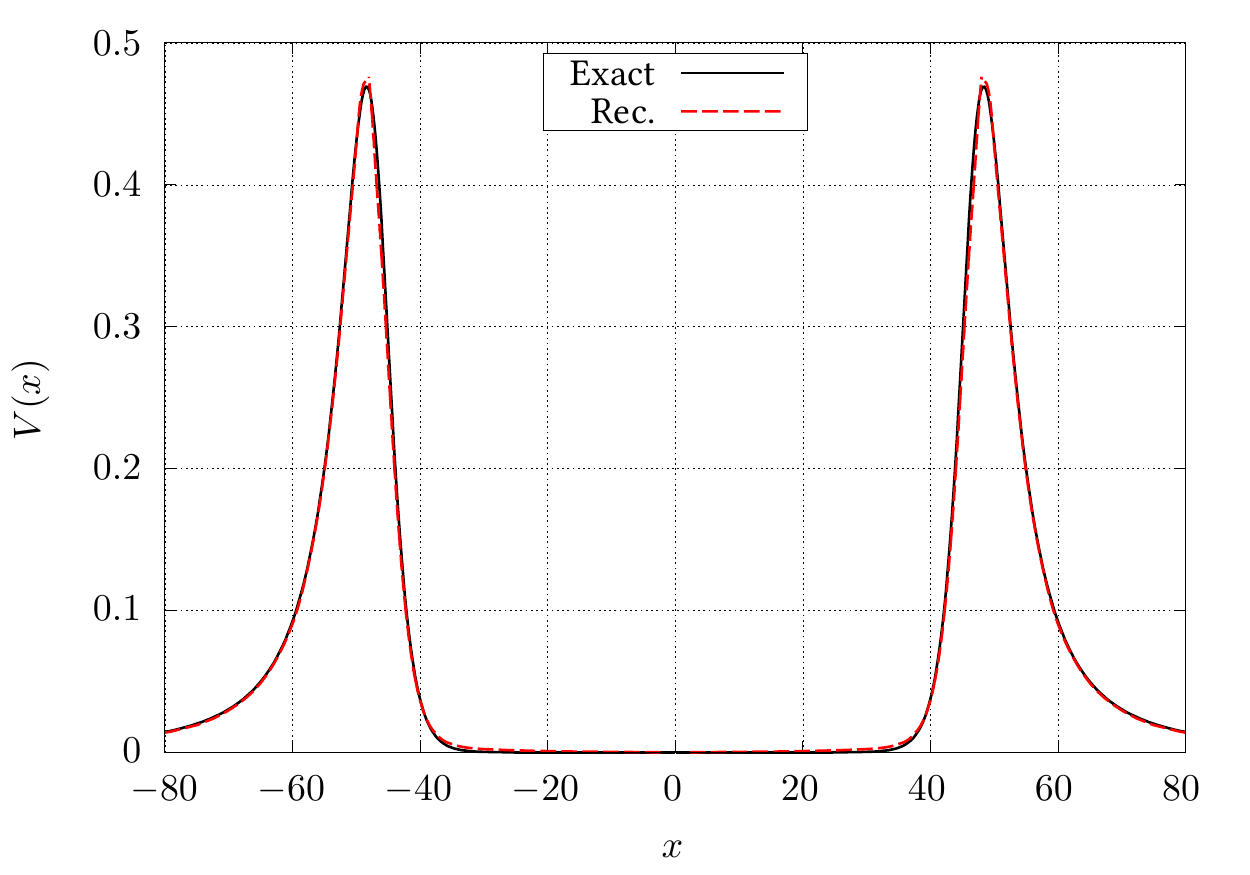}
	\end{minipage}
	\quad
	\begin{minipage}{0.45 \textwidth}
	\includegraphics[width=1.0\textwidth]{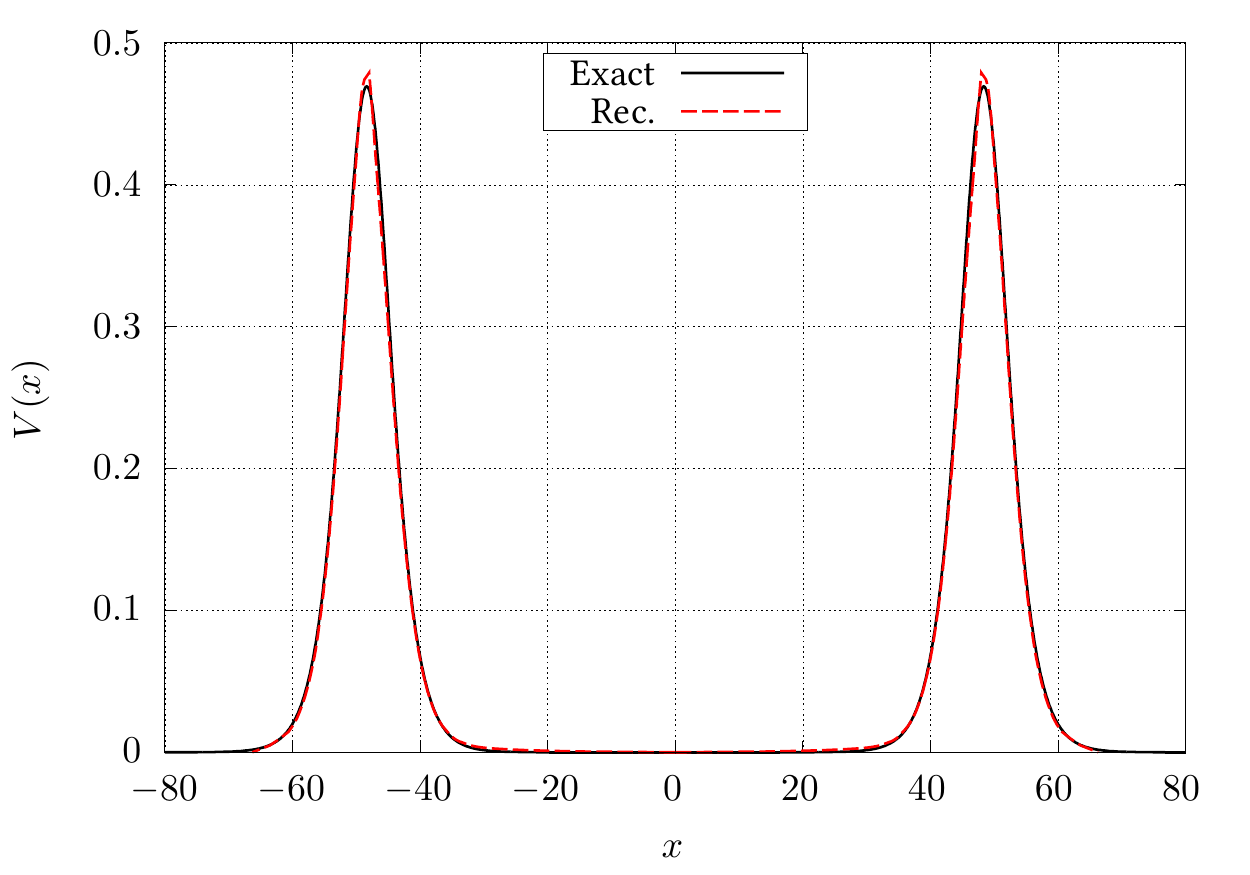}
	\end{minipage}
	\caption{In this figure we compare the exact DS potential and double PT potential (black solid) with our results for the reconstructed potential (red dashed) for $\lambda=10^{-5}$ and $l=3$. There are 20 trapped modes in both cases. \textbf{Left panel:} Here we show our result for the DS potential.
	\textbf{Right panel:} Here we show our result for the double PT potential.\label{DSPT_REC}}
\end{figure}

Here we provide additional results for reconstructed potentials for $\lambda=10^{-5}$ and $l=3$. The reconstruction of the DS and double PT potentials are shown in Fig. \ref{DSPT_REC}.
%
%
\subsection{Tabulated QNMs}\label{appendix_tables}
In the subsequent tables \ref{Table_DS1}, \ref{Table_DS2} and \ref{Table_PT} we provide the tabulated QNMs that we used in the inverse method. The numerical values of the DS potential were produced by a modified numerical code first introduced in \cite{1994MNRAS.268.1015K}, while the spectrum of the double P\"oschl-Teller potential can be obtained by solving an implicit equation as shown in \cite{Bueno:2017hyj}. 
\begin{table}[H]
	\centering
	\caption{The trapped QNM spectrum for the DS potential for different values of $l$ and $\lambda=10^{-5}$.
		\label{Table_DS1}}
	\begin{tabular}{|c| ll| ll| ll|}
		\hline
		& \multicolumn{2}{|c|}{$l=1$} & \multicolumn{2}{|c|}{$l=2$} & \multicolumn{2}{|c|}{$l=3$} \\
			\hline
			n & \text{Re}($\omega_n$) & \text{Im}($\omega_n$) & \text{Re}($\omega_n$) & \text{Im}($\omega_n$)& \text{Re}($\omega_n$) & \text{Im}($\omega_n$)  \\
			\hline
0	&	0.0350	&	1.96e-08	&	0.0367	&	6.44e-13	&	0.0378	&	6.70e-18	\\
1	&	0.0698	&	4.44e-07	&	0.0731	&	5.26e-11	&	0.0755	&	2.78e-15	\\
2	&	0.1043	&	3.30e-06	&	0.1092	&	9.03e-10	&	0.1127	&	1.10e-13	\\
3	&	0.1382	&	1.62e-05	&	0.1449	&	8.05e-09	&	0.1496	&	1.72e-12	\\
4	&	0.1715	&	6.28e-05	&	0.1801	&	4.99e-08	&	0.1859	&	1.64e-11	\\
5	&	0.2041	&	2.08e-04	&	0.2148	&	2.47e-07	&	0.2218	&	1.18e-10	\\
6	&	0.2358	&	6.01e-04	&	0.2490	&	1.06e-06	&	0.2572	&	6.92e-10	\\
7	&	0.2667	&	1.52e-03	&	0.2827	&	4.07e-06	&	0.2922	&	3.51e-09	\\
8	&	0.2968	&	3.30e-03	&	0.3159	&	1.43e-05	&	0.3267	&	1.58e-08	\\
9	&		&		&	0.3485	&	4.64e-05	&	0.3609	&	6.47e-08	\\
10	&		&		&	0.3805	&	1.41e-04	&	0.3947	&	2.48e-07	\\
11	&		&		&	0.4118	&	3.95e-04	&	0.4280	&	8.90e-07	\\
12	&		&		&	0.4423	&	1.01e-03	&	0.4610	&	3.02e-06	\\
13	&		&		&	0.4721	&	2.30e-03	&	0.4936	&	9.77e-06	\\
14	&		&		&		&		&	0.5257	&	3.02e-05	\\
15	&		&		&		&		&	0.5573	&	8.88e-05	\\
16	&		&		&		&		&	0.5884	&	2.48e-04	\\
17	&		&		&		&		&	0.6187	&	6.45e-04	\\
18	&		&		&		&		&	0.6483	&	1.53e-03	\\
19	&		&		&		&		&	0.6773	&	3.22e-03	\\
			\hline 
		\end{tabular}
	\end{table}
\begin{table}[H]
	\centering
	\caption{The trapped QNM spectrum for the DS potential for different values of $l$ and $\lambda=10^{-4}$.
		\label{Table_DS2}}
	\begin{tabular}{|c| ll| ll| ll|}
		\hline
		& \multicolumn{2}{|c|}{$l=1$} & \multicolumn{2}{|c|}{$l=2$} & \multicolumn{2}{|c|}{$l=3$} \\
			\hline
			n & \text{Re}($\omega_n$) & \text{Im}($\omega_n$) & \text{Re}($\omega_n$) & \text{Im}($\omega_n$)& \text{Re}($\omega_n$) & \text{Im}($\omega_n$)  \\
			\hline
0	&	0.0441	&	6.66e-08	&	0.0467	&	3.38e-12	&	0.0486	&	9.13e-17	\\
1	&	0.0877	&	1.69e-06	&	0.0929	&	3.55e-10	&	0.0967	&	3.31e-14	\\
2	&	0.1304	&	1.44e-05	&	0.1384	&	7.04e-09	&	0.1440	&	1.47e-12	\\
3	&	0.1719	&	7.92e-05	&	0.1828	&	7.15e-08	&	0.1903	&	2.70e-11	\\
4	&	0.2121	&	3.37e-04	&	0.2263	&	5.10e-07	&	0.2355	&	3.03e-10	\\
5	&	0.2506	&	1.16e-03	&	0.2687	&	2.92e-06	&	0.2799	&	2.52e-09	\\
6	&	0.2876	&	3.22e-03	&	0.3102	&	1.43e-05	&	0.3234	&	1.72e-08	\\
7	&		&		&	0.3506	&	6.16e-05	&	0.3662	&	1.00e-07	\\
8	&		&		&	0.3898	&	2.36e-04	&	0.4082	&	5.21e-07	\\
9	&		&		&	0.4277	&	7.94e-04	&	0.4494	&	2.45e-06	\\
10	&		&		&	0.4642	&	2.26e-03	&	0.4898	&	1.06e-05	\\
11	&		&		&		&		&	0.5295	&	4.23e-05	\\
12	&		&		&		&		&	0.5681	&	1.56e-04	\\
13	&		&		&		&		&	0.6056	&	5.24e-04	\\
14	&		&		&		&		&	0.6418	&	1.54e-03	\\
15	&		&		&		&		&	0.6769	&	3.84e-03	\\
			\hline 
		\end{tabular}
	\end{table}
\begin{table}[H]
	\centering
	\caption{The trapped QNM spectrum for the double P\"oschl-Teller potential for different values of $l$ and $\lambda=10^{-5}$.
		\label{Table_PT}}
	\begin{tabular}{|c| ll| ll| ll|}
		\hline
		& \multicolumn{2}{|c|}{$l=1$} & \multicolumn{2}{|c|}{$l=2$} & \multicolumn{2}{|c|}{$l=3$} \\
			\hline
			n & \text{Re}($\omega_n$) & \text{Im}($\omega_n$) & \text{Re}($\omega_n$) & \text{Im}($\omega_n$)& \text{Re}($\omega_n$) & \text{Im}($\omega_n$)  \\
			\hline
0	&	0.0362	&	7.28e-07	&	0.0385	&	1.70e-09	&	0.0402	&	3.67e-12	\\
1	&	0.0720	&	3.83e-06	&	0.0766	&	9.46e-09	&	0.0799	&	2.10e-11	\\
2	&	0.1071	&	1.31e-05	&	0.1139	&	3.49e-08	&	0.1189	&	8.12e-11	\\
3	&	0.1415	&	3.91e-05	&	0.1505	&	1.14e-07	&	0.1569	&	2.80e-10	\\
4	&	0.1751	&	1.09e-04	&	0.1863	&	3.57e-07	&	0.1941	&	9.23e-10	\\
5	&	0.2077	&	2.91e-04	&	0.2213	&	1.08e-06	&	0.2306	&	2.96e-09	\\
6	&	0.2394	&	7.26e-04	&	0.2556	&	3.19e-06	&	0.2664	&	9.26e-09	\\
7	&	0.2701	&	1.66e-03	&	0.2893	&	9.22e-06	&	0.3016	&	2.85e-08	\\
8	&	0.3000	&	3.41e-03	&	0.3224	&	2.61e-05	&	0.3363	&	8.60e-08	\\
9	&		&		&	0.3547	&	7.18e-05	&	0.3704	&	2.56e-07	\\
10	&		&		&	0.3864	&	1.91e-04	&	0.4041	&	7.50e-07	\\
11	&		&		&	0.4173	&	4.87e-04	&	0.4373	&	2.17e-06	\\
12	&		&		&	0.4474	&	1.16e-03	&	0.4700	&	6.15e-06	\\
13	&		&		&	0.4767	&	2.49e-03	&	0.5022	&	1.72e-05	\\
14	&		&		&		&		&	0.5340	&	4.70e-05	\\
15	&		&		&		&		&	0.5651	&	1.25e-04	\\
16	&		&		&		&		&	0.5957	&	3.22e-04	\\
17	&		&		&		&		&	0.6255	&	7.84e-04	\\
18	&		&		&		&		&	0.6545	&	1.76e-03	\\
19	&		&		&		&		&	0.6830	&	3.53e-03	\\
			\hline 
		\end{tabular}
	\end{table}
\subsection{Reconstructed Throat Parameter $\lambda$}\label{appendix_throat}
In table \ref{Table_throat} we provide the tabulated values for the reconstruction of the throat parameter $\lambda$ as described in Sec. \ref{Throat}.
The relative error is defined as $\Delta_\lambda\equiv |\lambda-\lambda_\text{rec.}|/\lambda$ and provided in the last row. 
\begin{table}[H]
	\centering
	\caption{The table shows the reconstructed values $\lambda_\text{rec.}$ of $\lambda=($1e-4, 1e-5$)$ obtained by relating the width of the true potential well at $E_\text{max}/2$ with the reconstructed one for $l=(1,2,3)$. The corresponding relative error $\Delta_\lambda$ is given in the last row. 
		\label{Table_throat}}
	\begin{tabular}{|c| ccc| ccc| }
		\hline
		$\lambda $& \multicolumn{3}{|c|}{1e-4} & \multicolumn{3}{|c|}{1e-5}  \\
			\hline
		$l$		&	$1$	&	$2$	&	$3$	&	$1$	&	$2$	&	$3$	\\
			\hline
$\lambda_\text{rec.}$		&	0.863e-4	&	0.944e-4	&	0.960e-4	&	0.840e-5	&	0.929e-5	&	0.951e-5		\\
$\Delta_\lambda $			&	0.137		&	0.056		&	0.040		&	0.160	&	0.071	&	0.049\\
			\hline 
		\end{tabular}
	\end{table}
\end{document}